\documentclass[letterpaper, 10 pt, conference]{ieeeconf}  

\IEEEoverridecommandlockouts                              

\usepackage{amsfonts}
\usepackage{mathtools}

\usepackage{bigints}
\usepackage{framed}
\usepackage{dirtytalk}
\usepackage{color}

\usepackage{gensymb}
\usepackage{accents}

\usepackage{booktabs}

\title{\LARGE \bf
Computation of Centroidal Voronoi Tessellations in High Dimensional spaces
}

\author{Bhagyashri Telsang$^{1}$ and Seddik M Djouadi$^{1}$
\thanks{$^{1}$Bhagyashri Telsang and Seddik Djouadi are with the Department of Electrical Engineering and Computer Science,
        University of Tennessee Knoxville, USA
        {\tt\small \{btelsang,mdjouadi\}@utk.edu}}%
}

\begin{document}

\maketitle
\thispagestyle{empty}
\pagestyle{empty}

\begin{abstract}
Owing to the natural interpretation and various desirable mathematical properties, centroidal Voronoi tessellations (CVT) have found a wide range of applications and correspondingly a vast development in their literature. However the computation of CVT in higher dimensional spaces still remains difficult. In this paper, we exploit the non-uniqueness of CVTs in higher dimensional spaces for their computation. We construct such high dimensional tessellations from CVTs in one-dimensional spaces. We then prove that such a tessellation is centroidal under the condition of independence among densities over the one-dimensional spaces considered. Various numerical evaluations backup the theoretical result through the low energy of the tessellations. The resulting grid-like tessellations are obtained efficiently with minimal computation time. 

\end{abstract}


\section{Introduction}
\label{sec::introduction}

Voronoi diagram is a partition of a set into subsets containing elements that are close to each other according to a certain metric. Even though they date centuries, Voronoi tessellations have been found immensely helpful in various applications ranging from health to computer graphics to natural sciences. The first documented application of Voronoi tessellations appeared in \cite{JohnSnowCholera} on the 1854 cholera epidemic in London in which it is demonstrated that proximity to a particular well was strongly correlated to deaths due to the disease \cite{LiliJu2011}. In more recent decades, Voronoi tessellations have almost become a common basis tool for path planning algorithms by multi-robot systems in the field of coverage control \cite{CortesMartinezMobileSensing}. 

In line with the popularity of Centroidal Voronoi tessellations (CVT), remarkable amount of contributions have been made to further their development. \cite{ConstrainedCVT_QiangDu} refines the notion of Constrained CVTs and derives various properties like their characterization as energy minimizers. Focusing on one-dimensional Voronoi diagrams, \cite{OneDimVD_Franz} develops an optimal algorithm for computing collinear weighted Voronoi diagrams that is conceptually simple and attractive for practical implementations. \cite{InverseVornoiGoberna} studies the inverse Voronoi problem in-depth. 

Despite the wide applicability and vast development in the literature pertaining to CVTs, there remain challenges and open questions, especially in high dimensional spaces. For dimensions greater than one, rigorously verifying a given centroidal Voronoi tessellation is a local minimum can prove difficult, for example \cite{UniquenessCVT_Urschel} uses variational techniques to give a full characterization of the second variation of a centroidal Voronoi tessellation and provides sufficient conditions for a centroidal Voronoi tessellation to be a local minimum. Moreover, in high dimensional spaces, the number of CVTs under certain conditions and their quality is elusive, and their computation remains difficult.
 
In this paper, we show how one can employ CVTs in one-dimensional spaces to construct a tessellation in a higher dimensional space. Then we prove that such a tessellation is a CVT in the higher dimensional space under certain conditions. Such a construction is a simple and yet a powerful technique that is guaranteed to render at least one of the many non-unique CVTs in high dimensions in a fast and efficient way with minimal computational requirements. 

The desired number of centroids in the higher dimensional space is a product of the number of centroids in the one-dimensional spaces. Accordingly, if the dimension is too large, one can limit the number of centroids in one-dimensional spaces in order to keep the total number of centroids meaningful. Doing so also has the advantages of further reduction in computational time. 

The paper is structured as follows. In Section \ref{sec::preliminaries} we formally define CVTs, discuss their uniqueness properties, and existing methods to compute them. In Section \ref{sec:ComputationCVTHigherDimensions} we construct a tessellation in a higher dimensional space and prove that it is also a CVT. In Section \ref{sec::numericalresults} we provide numerical results of CVTs under different high dimensional spaces, densities and number of centroids. Finally, we present some conclusions and remarks on future work in Section \ref{sec::conclusion}.

\section{Preliminaries}
\label{sec::preliminaries}

Consider a region $\Omega \subset \mathbb{R}^n, \ n \geq 1$. Let $N \in \mathbb{N}$, $V_i \subset \Omega, \ \forall i \in I_N$, and denote index set as $I_N = \{1, 2, \hdots, N\}$. Let $\rho(.)$ denote a measure of information or the probability density over $\Omega$. 

\begin{itemize}
    \item[1] Tessellation: $\{V_i\}_{i\in I_N}$ is a tessellation of $\Omega$ if $V_i \cap V_j = \emptyset$ for $i \neq j$, and $\cup_{i\in I} {V}_i = {\Omega}$. 
    \item[2] Voronoi region and generators: The Voronoi region ${V}_{z_i}$ of the Voronoi generator $z_i$ is ${V}_{z_i} = \{x \in \Omega : ||x - z_i|| < ||x-z_j||, \ i \neq j \ \text{and} \ i,j \in I_N \}$. 
    \item[3] Voronoi tessellation: The set of Voronoi regions $\textbf{V}_{\textbf{z}} = \{V_{z_i}\}_{i \in I_N}$ of $\{z_i\}_{i \in I_N}$ is called a Voronoi tessellation $\{ \textbf{z},\textbf{V}_{\textbf{z}}\}$.
\end{itemize}

The mass centroid of a region $V_i \subset \Omega$ under the probability density function $\rho(.)$ is defined as:

\begin{equation}
    z_{V_i,\rho}^c  = \frac{\int_{V_i} x \rho(x) dx}{\int_{V_i}\rho(x) dx}
    \label{eq:MassCentroidDefn}
\end{equation}

\begin{figure}
    \centering
    \includegraphics[width=\columnwidth]{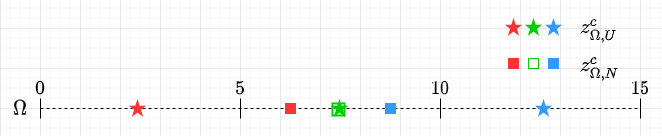}
    \caption{Centroidal Voronoi Tessellations of $[0,15]$ under Uniform and Normal distributions, denoted in star and square symbols respectively.}
    \label{fig:CVT_defn}
\end{figure}

\noindent A Voronoi tessellation in which the generators are the mass centroids of their respective Voronoi regions is called a \textit{Centroidal Voronoi Tessellation} (CVT), \cite{CVT_QiangDu}. The CVT obtained for 3 generators in the region $\Omega = [0,15]$ under Uniform and Normal distributions -- $\mathcal{U}(0,15)$ and $\mathcal{N}(7.5,1)$ -- are shown in Fig. \ref{fig:CVT_defn}. The generators under Uniform ad Normal distribution over $\Omega$ are marked in star and square symbols respectively.

\subsection{Uniqueness of CVT}
\label{subsec:UniquenessofCVT}

Given a region $\Omega \subset \mathbb{R}^n$, a positive integer $N$, and a density function $\rho(x)$ on $\Omega$, consider the functional $\mathcal{F}$ with any $N$ points $\{z_i\}_{i \in I_N} \in \Omega$ and any tessellation $\{V_i\}_{i \in I_N}$ of $\Omega$ as its input arguments:

\begin{equation}
	\mathcal{F}((z_i,V_i), i \in I_N) = \sum_{i \in I_N} \int_{x\in V_i} \rho(x) ||x-z_i||^2 dx
	\label{eq:functionalF_QiangDu}
\end{equation}

\noindent Proposition $3.1$ in \cite{CVT_QiangDu} states that a necessary condition for the function $\mathcal{F}$ to be minimized is that $\{V_i\}_{i \in I_N}$ are the Voronoi regions corresponding to $\{z_i\}_{i \in I_N}$, and simultaneously, $\{z_i\}_{i \in I_N}$ are the centroids of their respective Voronoi regions. In other words, the minimizer of $\mathcal{F}$ is a Centroidal Voronoi Tessellation. 

Additionally, if the tessellation in \eqref{eq:functionalF_QiangDu} is fixed to be the Voronoi tessellation of $\{z_i\}_{i \in I_N}$, then the following functional $\mathcal{K}$ has the same minimizer as $\mathcal{F}$, \cite{CVT_QiangDu}.  

\begin{equation}
	\mathcal{K}((z_i), i \in I_N) = \sum_{i \in I_N} \int_{x\in V_{z_i}} \rho(x) ||x-z_i||^2 dx
	\label{eq:functionalK_QiangDu}
\end{equation}

\noindent This functional $\mathcal{K}$ is also referred to as the energy of the tessellation or the quantization energy. It is stated and proved in Lemma $3.4$ in \cite{CVT_QiangDu} that $\mathcal{K}$ is continuous and that it possesses a global minimum. Moving from existence of the CVT to its uniqueness, \cite{CVT_QiangDu} also mentions that $\mathcal{K}$ may have local minimizers. 

It is showed in \cite{CVT_Fleischer} that the solution of \eqref{eq:functionalK_QiangDu} is unique in one-dimensional regions with a logarithmically concave continuous probability density function of finite second moment. As reiterated in \cite{UniquenessCVT_Urschel}, for $n=1$, the logarithmic concavity condition implies that any CVT is a local minimum, and further, that there is a unique CVT that is both a local and a global minimum of $\mathcal{K}(\textbf{z})$, where $\textbf{z} = \{z_i\}_{i\in I_N}$. Accordingly, since the two distributions -- Uniform and Normal -- considered in Fig. \ref{fig:CVT_defn} on the one-dimensional region $\Omega = [0,15]$ are log-concave with finite second moment, we have that the CVTs showed therein are the global minima for the two distributions. 

The solution of \eqref{eq:functionalK_QiangDu} is also called scalar quantizer for $n=1$ and vector quantizer for higher dimensions. The conditions for uniqueness of vector quantizers for the general case, that is, without any assumptions on the region, density or the number of quantizers $N$, remains an open area of research. However, it is proved in \cite{UniquenessCVT_Urschel} that for $N=2$, there does not exist a unique CVT for any density for $n > 1$.

\begin{figure}
    \centering
    \includegraphics[width=0.9\columnwidth]{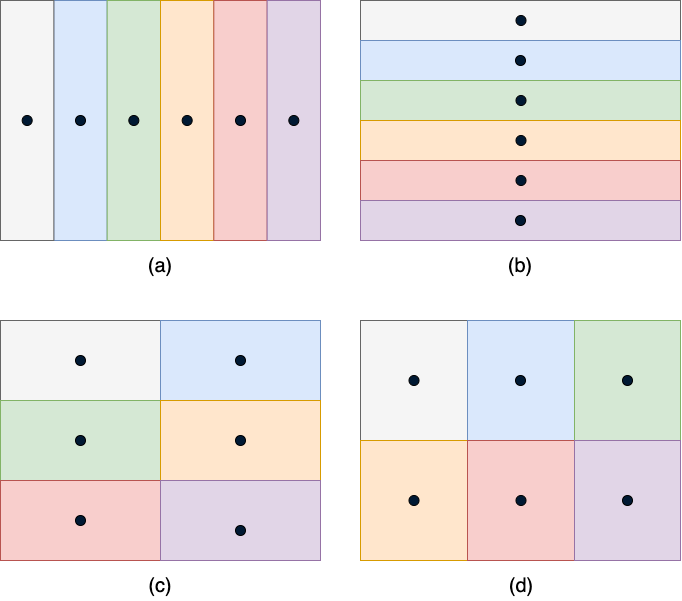}
    \caption{CVT is not necessarily unique in $\mathbb{R}^2$: Four centroidal Voronoi tessellations, each of six generators, in a rectangle in $\mathbb{R}^2$ with uniform density. Voronoi regions are shown in different colors while their centroids are marked as black dots.}
    \label{fig:NonUniqueCVT_example}
\end{figure}

For a graphic illustration on non-uniqueness of CVTs, consider a rectangle in $\mathbb{R}^2$ with six generator points under Uniform distribution. As shown in Fig. \ref{fig:NonUniqueCVT_example}, there are multiple CVTs: all the four Voronoi tessellations shown are centroidal, and additional CVTs can be obtained through rotations. 

To summarize, the solution of \eqref{eq:functionalK_QiangDu} is unique in one-dimensional regions for log-concave density functions but there does not exist a unique CVT for any density for dimensions greater than one \cite{UniquenessCVT_Urschel}. 

\subsection{Computation of CVT}
\label{sec::ComputeCVTmethods}

Given a region $\Omega \subset \mathbb{R}^n$, a number of generators $N$, and a density function $\rho(x)$ over $\Omega$, there are various iterative algorithms to compute a CVT in $\Omega$. As noted in the previous Section, in general, the CVT need not be unique for any dimensional region unless certain conditions are imposed on the density function. Accordingly, the solutions rendered by all the algorithms, to compute the CVT, need not be the unique global minimizers. In this Section, we first describe perhaps the most widely employed of all the algorithms: Lloyd's algorithm, and then show a computation method to obtain the exact solution of the problem of finding the CVT in one-dimensional regions.  

Introduced in \cite{CVT_Lloyd_original} to find the optimal quantization in pulse-code modulation, Lloyd's algorithm, also known as Voronoi iteration or relaxation, has been modified or adapted in various fields. At the core of it, Lloyd's algorithm is an iteration between constructing Voronoi tessellations and their centroids:


\begin{framed}
\noindent Given: $\Omega \subset \mathbb{R}^n$, $N$, $\rho(x)$ \\
Initialize: Generators $\textbf{z} = \{z_i\}_{i \in I}$, where each $z_i \in \Omega$
\begin{itemize}
	\item[1] Construct the Voronoi tessellation $\textbf{V}_\textbf{z}$.
	\item[2] Compute the mass centroids $z^c_{\textbf{V}_{\textbf{z},\rho}}$ of $\textbf{V}_\textbf{z}$. 
	\item[3] If the computed centroids meet certain stopping criteria then terminate. If not, then set $\textbf{z} = z^c_{\textbf{V}_{\textbf{z},\rho}}$, and return to Step 1.
\end{itemize}
 \end{framed}

\noindent Since we are looking for a CVT, the search/iterations stop when the centroids of the Voronoi regions are the generators. Accordingly, the stopping criteria can be defined to have the generators same as the centroids with some tolerance. 

Even though Lloyd's algorithm is iterative and approximate, it has certain desirable convergence properties. \cite{ConvergenceLloyd_QiangDu} rigorously proves various global convergence properties of the Lloyd's algorithm and surveys all the results concerning the convergence of the Lloyd's algorithm. Specifically for one-dimensional spaces with log-concave density function, the local convergence using the Lloyd's algorithm has been proved in \cite{CVT_1d_uniquenessKieffer}. 


Another method to obtain a CVT in a one-dimensional region is SNLE (system of nonlinear equations). Perhaps the most straightforward way to compute a CVT, the core idea behind SNLE, which is the analytical computation of the CVT in one-dimensional region, is to parameterize the Voronoi regions in terms of their centroids:  $V_i = [ \frac{z_{i-1}+z_i}{2}, \frac{z_i+z_{i+1}}{2}] , \forall i \in I_N$. The mass centroids from \eqref{eq:MassCentroidDefn}, with  $z_{V_i,\rho}^c$ denoted as $z_i^c$ for ease of notation, can then be written in terms of the parameterized regions:

\begin{equation}
	z_{i}^c  = \frac{\int_\frac{z_{i-1}^c+z_i^c}{2}^{\frac{z_i^c+z_{i+1}^c}{2}} x \rho(x) dx}{\int_\frac{z_{i-1}^c+z_i^c}{2}^{\frac{z_i^c+z_{i+1}^c}{2}}\rho(x) dx}  
	\label{eq:Centroids_Parameterizing_1D}
\end{equation}

Writing \eqref{eq:Centroids_Parameterizing_1D} for all the centroids results in $N$ equations and $N$ number of unknowns: $z_i^c$'s. This is the system of nonlinear equations (SNLE), whose solution is the set of centroids of the CVT. In one-dimensional regions, the centroids are directly used to define the corresponding Voronoi partitions.

While the Lloyd's algorithm and SNLE obtain the CVT in a deterministic way, MacQueen's method developed in \cite{JMacQueen} takes a probabilistic approach. It was developed to partition an $n$-dimensional population into $k$ sets on the basis of a sample, and the process is also called $k-$means. As noted in \cite{CVTQiangDuParallel}, despite the elegance of MacQueen's method, the algorithm can be slow to converge due to a single random sampling point. Accordingly, \cite{CVTQiangDuParallel} presents a new method that can be viewed as a probabilistic version of Lloyd’s method and/or a generalization of the random MacQueen’s method. Some other algorithms that have faster convergence than Lloyd's have been proposed in \cite{CVT_LlyodsAlt_FastConvergenceLiu}, \cite{CVT_LloydsAlt_FastConvergenceWang}, \cite{CVT_LlyodsAlt_FastConvergenceHateley}.

However, the computation of CVT in higher dimensional spaces still remains a challenge. \cite{CVTQiangDuParallel} considers parallel implementations of Lloyd's, MacQueen's and a method developed therein, and presents the CVT results for regions up to two dimensions while noting that the method can be extended to higher dimensions. Although \cite{CVT_LlyodsAlt_FastConvergenceHateley} results in faster convergence to a CVT, it does not easily extend to higher dimensions. In the next Section, we present a way to extend SNLE (or any method to obtain one-dimensional CVTs) to obtain a higher dimensional CVT.

\section{COMPUTATION OF CVT IN HIGHER DIMENSIONS}
\label{sec:ComputationCVTHigherDimensions}

In this Section, we first propose a simple method to obtain a tessellation in any higher dimensional space from CVTs in one-dimensional spaces. Then, we present the proof that a tessellation constructed in such a way is also a CVT. 

Consider a region $\Omega \subset \mathbb{R}^n, \ n > 1$, and $\Omega = \Omega_1 \times \Omega_2 \times \hdots \times \Omega_n$. Let $\rho(.)$ be the probability density function over $\Omega$, and $\rho_i(.)$ be the density function over $\Omega_i$, $\forall i \in I_n = \{1,2,\hdots, n\}$. That is, $\rho(.)$ is the joint density, and $\{\rho_i(.)\}_{i \in I_n}$ are the marginal densities. 

Let the number of generators in a CVT of $\Omega_i$ under $\rho_i(.)$ be $N_i$, and let the number of generators in a CVT of $\Omega$ under $\rho(.)$ be $N$, where $N = N_1 \times \hdots \times N_n$. Denote a CVT in $\Omega_i$ as $\{\textbf{z}^*_i, \textbf{V}_{\textbf{z}^*_i}\}$. Here, $\textbf{z}^*_i = \{z^*_{i,j}\}_{j \in I_{N_i}}$ is the set of all the centroids of the CVT in $\Omega_i$, and $\textbf{V}_{\textbf{z}^*_i} = \{V_{z^*_{i,j}}\}_{j \in I_{N_i}}$ is the set of their respective Voronoi regions. Similarly, denote a CVT in $\Omega$ as $\{\textbf{z}^*,\textbf{V}_{\textbf{z}^*}\}$, where $\textbf{z}^* = \{z_k^*\}_{k \in I_N}$ denotes the centroids, $\textbf{V}_{\textbf{z}^*} = \{V_{z^*_k}\}_{k \in I_N}$  denotes their corresponding Voronoi regions.

Note that $\textbf{z}^* \in \mathbb{R}^N$ while $\textbf{z}^*_i \in \mathbb{R}^{N_i} $, and $z^*_k \in \mathbb{R}^n$ while $z^*_{i,j} \in \mathbb{R}$. Denote each element of $z^*_k \in \mathbb{R}^n$ by indexing as $z^*_k(1), z^*_k(2), \hdots, z^*_k(n)$. Additionally, note that $V_{z^*_k} \subset \mathbb{R}^n$ while $V_{z^*_{i,j}} \subset \mathbb{R}$.

Let $I_{n\times N}$ denote the matrix containing all combinations of vectors $I_{N_i}, \forall i \in I_n$. For example, if $n=2, N_1 = 2, N_2 = 3,$ then $N = 2\times3=6$ and

\begin{equation}
I_{n\times N} = 
\begin{bmatrix}
1 & 1 & 1 & 2 & 2 & 2 \\
1 & 2 & 3 & 1 & 2 & 3
\end{bmatrix}
\end{equation}

With all the notations defined, we now present a straightforward method of constructing a tessellation in $\Omega$ from CVTs in $\Omega_i's$:

\begin{framed}

\hspace{1cm} \textbf{Tessellation construction in $\Omega$}

\noindent For each dimension $i \in I_n$, construct a CVT in $\Omega_i$: $\{\textbf{z}^*_i, \textbf{V}_{\textbf{z}^*_i}\}$

\noindent Obtain the $n$ coordinates of each centroid in $\Omega$ and its Voronoi region as:

$\forall k \in I_N:$

\hspace{0.5cm}$\forall i \in I_n:$ 

\hspace{1cm} $z^*_k =  z^*_{ I_{n \times N}(i,k)}$ 

\hspace{1cm} $V_{z^*_k}=  V_{z^*_{ I_{n \times N}(i,k)}}$ 

\noindent The set of all the centroids $\{z_k^*\}_{k \in I_N}$ and their Voronoi regions $\{V_{z^*_k}\}_{k \in I_N}$ make the tessellation in $\Omega$: $\{\textbf{z}^*,\textbf{V}_{\textbf{z}^*}\}$
\end{framed}

Having obtained the tessellation, we show in the following theorem that $\{\textbf{z}^*,\textbf{V}_{\textbf{z}^*}\}$ constructed from $\{\textbf{z}^*_i, \textbf{V}_{\textbf{z}^*_i}\}_{i \in I_n}$ is a CVT in $\Omega$.

{\textbf{Theorem:}} If events in $\Omega_i$ are independent of those in $\Omega_j$, $\forall i \neq j, \ i,j \in I_n$, then for some $k_i \in I_{N_i}, \forall i \in I_n$ and $\forall k \in I_N,$ we have:

\begin{equation}
z^*_k = (z^*_{1,k_1}, \ z^*_{2,k_2}, \hdots, z^*_{n,k_n})
\label{eq:TheoremCentroidStatement}
\end{equation}

\begin{equation}
V_{z^*_k} = V_{z^*_{1,k_1}} \times  V_{z^*_{2,k_2}} \times \hdots \times  V_{z^*_{n,k_n}}
\label{eq:TheoremVoronoiStatement}
\end{equation}

{\textbf{Proof:}} 

Consider $x \in V_{z^*_{1,k_1}} \times \hdots \times V_{z^*_{n,k_n}}$, $\forall i \in I_n$, and denote $x = (x_1, \hdots, x_n)$. Because $V_{z^*_{i,k_i}}$ is the Voronoi region of $z^*_{i,k_i}$, $\forall i \in I_n$, we have for any $j_i \in I_{N_i}$:

\begin{align}
||x_i - z^*_{i,k_i}||_2 \leq ||x_i - z^*_{i,j_i}||_2 \nonumber \\
\implies (x_i - z^*_{i,k_i})^2 \leq (x_i - z^*_{i,j_i})^2
\label{eq:VoronoiRegionRnProof1}
\end{align}

Summing \eqref{eq:VoronoiRegionRnProof1} $\forall i \in I_n$,

\begin{align}
(x_1 - z^*_{1,k_1})^2 + & \hdots + (x_n - z^*_{n,k_n})^2 \leq \\   \ \ (x_1 & - z^*_{1,j_1})^2 + \hdots + (x_n - z^*_{n,j_n})^2  \\
\implies \sqrt((x_1 - z^*_{1,k_1})^2 &+ \hdots + (x_n - z^*_{n,k_n})^2) \leq \\ \sqrt((x_1 & - z^*_{1,j_1})^2 + \hdots + (x_n - z^*_{n,j_n})^2) \label{eq:VoronoiRegionRnProof2}
\end{align}

Let $\hat{z}^*_k = (z^*_{1,k_1},  z^*_{2,k_2}, \hdots, z^*_{n,k_n})$, then \eqref{eq:VoronoiRegionRnProof2} results in:

\begin{align}
  ||x-\hat{z}^*_k||_2 \ \ \leq \ \  ||x-z^*_j|| \nonumber \\
\implies   V_{\hat{z}^*_k} = V_{z^*_{1,k_1}} \times  \hdots \times V_{z^*_{n,k_1}}
\label{eq:VoronoiRegionRnProof2}
\end{align}

\noindent That is, $V_{z^*_{1,k_1}} \times  \hdots \times V_{z^*_{n,k_1}}$ is the Voronoi region of $\hat{z}^*_k$. We are yet to prove that $\hat{z}^*_k$ is a centroid of a CVT in $\Omega$ under $\rho(.)$, that is, $z^*_k = \hat{z}^*_k$.

Consider the $i^{th}$ coordinate of $\hat{z}^*_k$. Since $z^*_{i,k_i}$ is the centroid of $V_{z^*_{i,k_i}}$, by definition of centroid, we have:

\begin{align}
  z^*_{i,k_i} &= \frac{\int_{V_{z^*_{i,k_i}}} x_i \rho_1(x_i)}{\int_{V_{z^*_{i,k_i}}} \rho_n(x_i) dx_i} \nonumber  \\
  &= \frac{\int_{V_{z^*_{1,k_1}}} \rho_1(x_1)dx_1}{\int_{V_{z^*_{1,k_1}}} \rho_1(x_1)dx_1} \times \hdots \times \frac{\int_{V_{z^*_{i,k_i}}} x_i\rho_i(x_i)dx_i}{\int_{V_{z^*_{i,k_i}}} \rho_i(x_i)dx_i} \nonumber \\ 
 &\hspace{2cm} \times \hdots \frac{\int_{V_{z^*_{n,k_n}}} \rho_n(x_n)dx_n}{\int_{V_{z^*_{n,k_n}}} \rho_n(x_n)dx_n} 
 \label{eq:MassCentroidRnProof1}
\end{align}

Because the events in $\Omega_i$ are independent of those in $\Omega_j$, $\forall i \neq j, \ i,j \in I_n$, we have $\rho(x_1, \hdots, x_n) = \rho_1(x_1), \hdots, \rho_n(x_n)$. Substituting this relation in \eqref{eq:MassCentroidRnProof1} implies: 

\begin{align*}
  z^*_{i,k_i} &= \frac{\int_{V_{z^*_{1,k_1}}} \hdots  \int_{V_{z^*_{n,k_n}}} x_i \rho(x_1, \hdots, x_n) dx_1 \hdots dx_n }{\int_{V_{z^*_{1,k_1}}} \hdots \int_{V_{z^*_{n,k}}} \rho(x_1, \hdots, x_n) dx_1 \hdots dx_n }
\end{align*}

\noindent which, by definition of mass centroids, is the $i^{th}$ coordinate of the $k^{th}$ of the $N$ centroids -- $z^*_k$ -- in $\Omega$ with density $\rho(.)$. That is, $z^*_k(i) = z^*_{i,k_i}$. Since this holds for all $i \in I_n$ coordinates, we have $z^*_k = \hat{z}^*_k = (z^*_{1,k_1},  z^*_{2,k_2}, \hdots, z^*_{n,k_n})$, and hence proving \eqref{eq:TheoremCentroidStatement}. On the other hand, since $V_{z^*_{1,k_1}} \times  \hdots \times V_{z^*_{n,k_1}}$ is the Voronoi region of $\hat{z}^*_k$ from \eqref{eq:VoronoiRegionRnProof2}, and $z^*_k = \hat{z}^*_k$, we have $V_{z^*_{1,k_1}} \times  \hdots \times V_{z^*_{n,k_1}}$ is the Voronoi region of $z^*_k$, hence proving \eqref{eq:TheoremCentroidStatement}. Since this holds for all $N$ centroids in $\Omega$, we have that $\textbf{z}^* = \{z^*_k\}_{k \in I_N}$ and $\textbf{V}_{\textbf{z}^*} = \{V_{z^*_k}\}_{k \in I_N}$ is a CVT in $\Omega$ with density $\rho(.)$.

$\hfill \square$

Recall the non-uniqueness of CVTs in $\mathbb{R}^2$ from Fig. \ref{fig:NonUniqueCVT_example}, that shows some of the CVTs for $N=6$ generators in a rectangle in $\mathbb{R}^2$ with uniform density. We construct the CVT in Fig. \ref{fig:NonUniqueCVT_example}(d) under different conditions.

Consider $\Omega = [0,20] \times [0,10]$ in $\mathbb{R}^2$ with density $\rho(.) \sim \mathcal{N}(\mu,\Sigma)$, where $\mu = (12,7)$ and $\Sigma=[4 \ \ 0; 0 \ \ 1]$. Denote the CVT as $\{\textbf{z}^*,\textbf{V}_{\textbf{z}^*}\}$, where $\textbf{z}^* = (z^*_1, \hdots, z^*_6)$, and $z^*_k \in \mathbb{R}^2, \forall k \in I_6 = \{1, \hdots, 6\}$. 

On the other hand, let $N_1 = 3$ and $N_2 = 2$. Consider the unique CVT in $\Omega_1 = [0,20]$ for $\rho_1(.) \sim \mathcal{N}(12,4)$, which is denoted $\{\textbf{z}_1^*,\textbf{V}_{\textbf{z}_1^*}\}$. Note $\textbf{z}_1^* = (z^*_{1,1}, z^*_{1,2}, z^*_{1,3})$, and $z^*_{1,j} \in \mathbb{R}, \forall j \in I_3$. Similarly, consider the unique CVT in $\Omega_2 = [0,10]$ for $\rho_1(.) \sim \mathcal{N}(7,1)$, which is denoted $\{\textbf{z}_2^*,\textbf{V}_{\textbf{z}_2^*}\}$. Note $\textbf{z}_2^* = (z^*_{2,1}, z^*_{1,2})$, and $z^*_{1,j} \in \mathbb{R}, \forall j \in I_2$. These generators are shown in Fig \ref{fig:Is1plus1equalto2_step1}: the region $\Omega_1$ and the CVT generators in it are showed in pink, and the region $\Omega_2$ and the CVT generators in it are showed in blue. While this method of obtaining a CVT in higher dimensions by employing a combination of CVTs in $\mathbb{R}$ does not result in every possible CVT of the higher dimension under the given conditions, we are guaranteed to obtain at least one of them in a straightforward manner with minimal computation.

\begin{figure}
\centering
    \includegraphics[width=0.9\columnwidth]{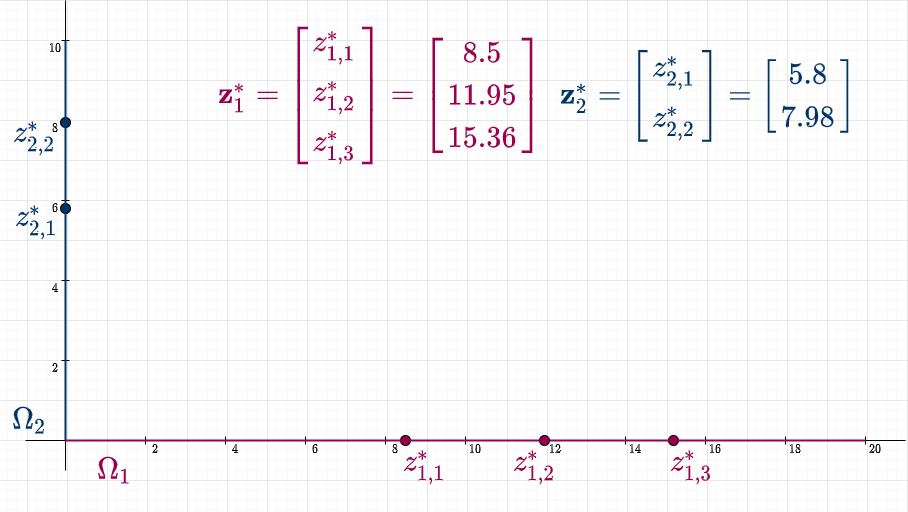}
    \caption{CVTs in $\Omega_1$ and $\Omega_2$ with densities $\mathcal{N}(12,4)$ and $\mathcal{N}(7,1)$ respectively.}
    \label{fig:Is1plus1equalto2_step1}
\end{figure}

\begin{figure}
\centering
    \includegraphics[width=0.9\columnwidth]{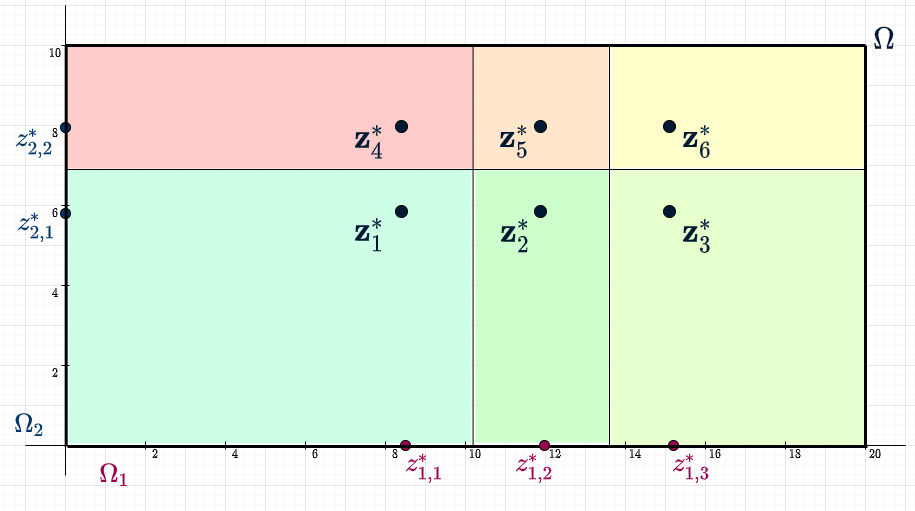}
    \caption{CVT in $\Omega$ with density $\mathcal{N}(\mu,\Sigma)$ where $\mu = [12; 7]$ and $\Sigma = [4 \ \ 0; 0 \ \ 1]$ .}
    \label{fig:Is1plus1equalto2_step1}
\end{figure}

\section{NUMERICAL RESULTS}
\label{sec::numericalresults} 

In this Section, we present a set of numerical results to demonstrate the ease of extension in higher dimensions through the time required to compute the CVT and the CVT energy. Additionally, for 2 and 3 dimensional spaces we also present the tessellations graphically. To obtain the CVTs in one-dimensional spaces, one can employ Lloyd's algorithm or solve the system of nonlinear equations; the latter being more desirable when $N_i's$ are low. 

\begin{figure}
    \centering
    \includegraphics[width=\columnwidth]{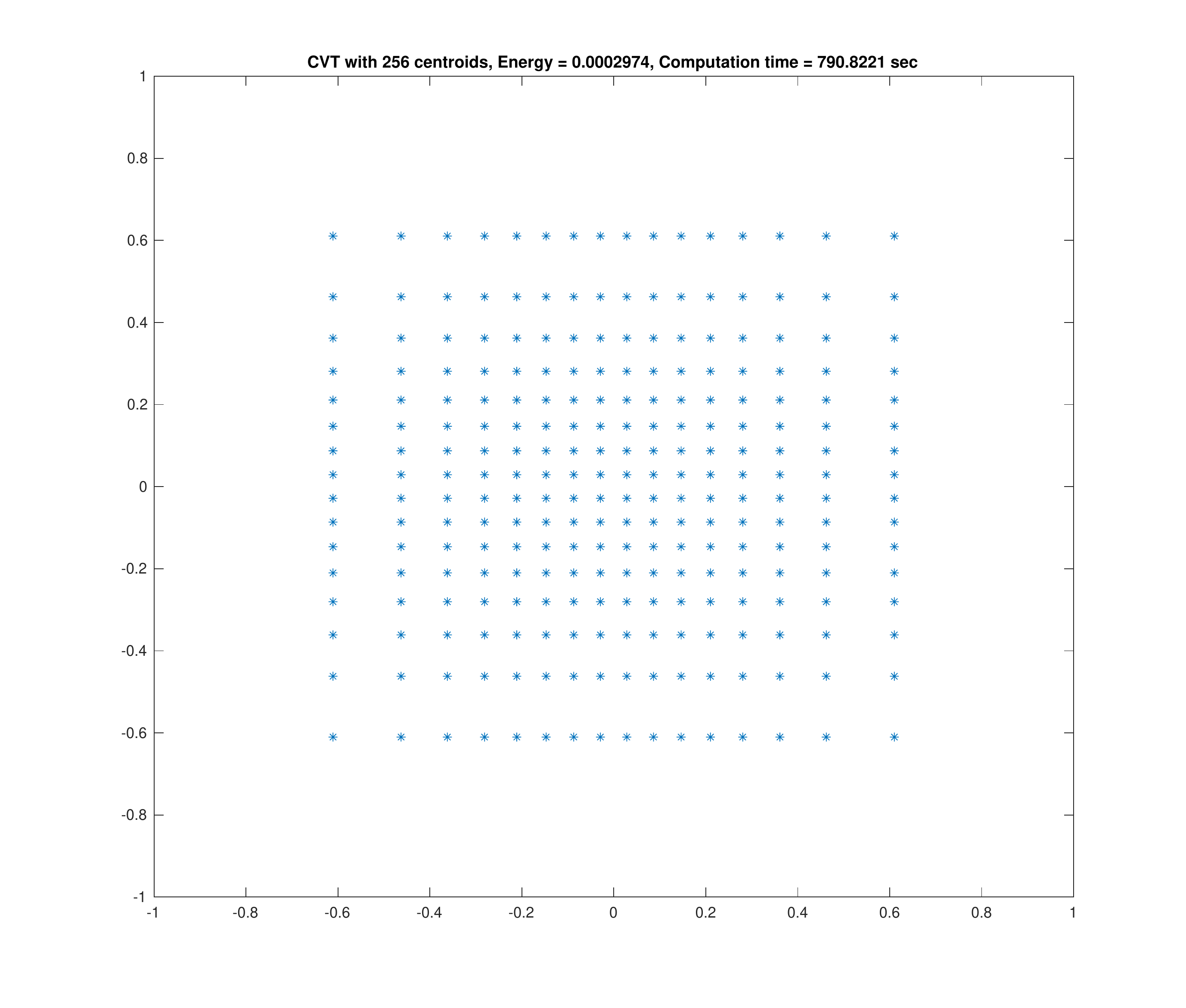}
    \caption{CVT of 256 centroids in a two-dimensional square under $\rho_1(x_1) = \rho_2(x_2) = e^{-10x^2}$ }
    \label{fig:CVT_2D_16x16_expminus10xsq}
\end{figure}

Consider the region $\Omega _1 = \Omega _2 = [-1, 1]$ and $\Omega = \Omega_1 \times \Omega_2 \subset \mathbb{R}^2$. Following the cases taken up in \cite{CVTQiangDuParallel}, we let the density function over $\Omega$ be $e^{-10x^2}$. The CVT of 256 centroids, obtained from $16$ in each dimension, in the aforementioned $\Omega$ is shown in Fig. \ref{fig:CVT_2D_16x16_expminus10xsq}. As designed and expected, the tessellation has a well drawn out grid-like structure with the intensity of centroids being higher in the center of $\Omega$. The resulting tessellation is a CVT with (low) energy of $2.9 \times 10^{-4}$ and was obtained in computational time as less as $13.18$ minutes in an ordinary laptop -- MacBook Air 2015 with 2.2 GHz Dual-Core Intel Core i7. 

\begin{figure}
    \centering
    \includegraphics[width=\columnwidth]{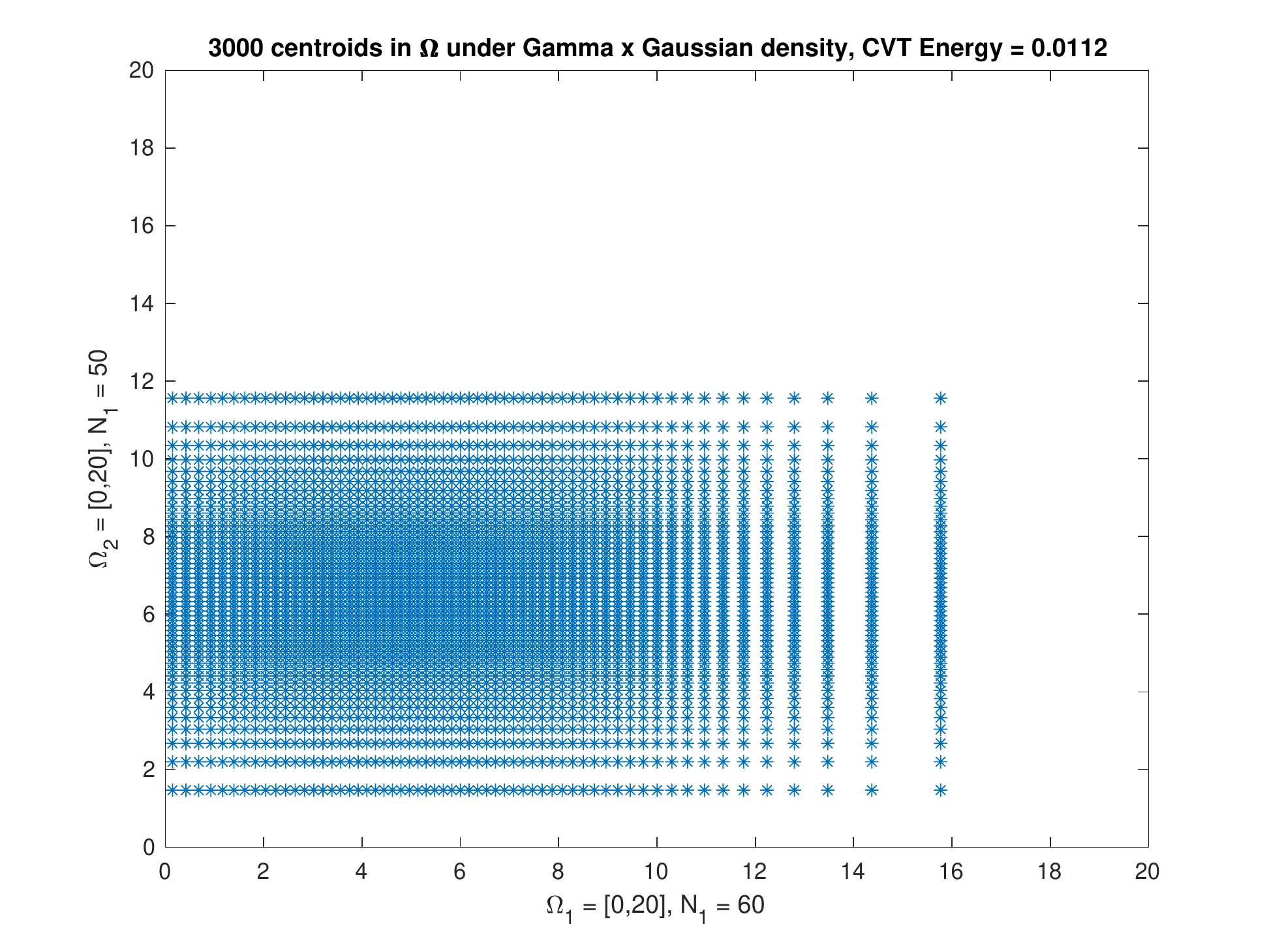}
    \caption{CVT of 3000 centroids in $\Omega = [0,20]\times [0,20]$ with Gaussian density.}
    \label{fig:CVT_2D_60x50_GammaGaussianWorkComp}
\end{figure}

Next case of demonstration is in the region $\Omega = [0,20] \times [0,20]$ with a Gaussian density of means $\begin{bmatrix} 5 \\ 6.5 \end{bmatrix}$ and variance $\begin{bmatrix} 2 \ \ 0 \\ 0 \ \ 1 \end{bmatrix}$. Here, we consider $N_1 = 60$ and $N_2 = 50$, implying total number of centroids in $\Omega$ being $N=3000$. The resulting CVT with energy $0.0112$ is shown in Fig. \ref{fig:CVT_2D_60x50_GammaGaussianWorkComp}. 

Moving to 3 dimensional spaces, we consider $\Omega = [0,10]\times [0,10] \times [0,10]$ under Gaussian density with mean $\mu$ and variance $\Sigma$ as:

\begin{equation*}
\mu =  \begin{bmatrix} 6 \\ 5 \\ 1 \end{bmatrix} \hspace{1cm}
\Sigma = \begin{bmatrix} 2 \ \ 0 \ \ 0 \\ 0 \ \ 1 \ \ 0 \\ 0 \ \ 0 \ \ 1 \end{bmatrix}
\end{equation*}

\begin{figure}
    \centering
    \includegraphics[width=0.9\columnwidth]{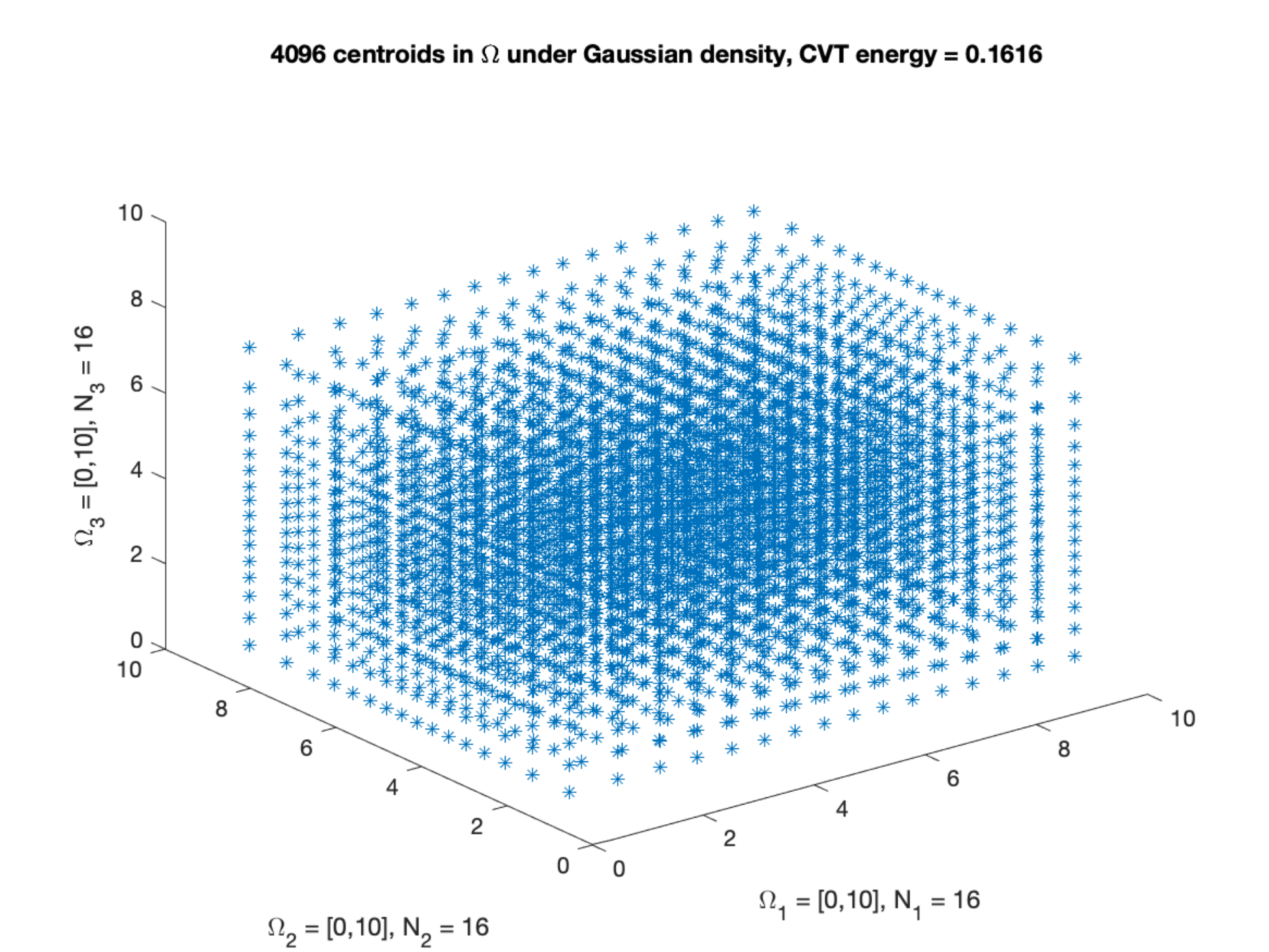}
    \caption{Scalability and generalizability to any density: CVT of 4096 centroids in a 3D space under Gaussian density.}
    \label{fig:CVT_3D_16x16x16_Gaussian62_51_3p51}
\end{figure}

\noindent The resulting well-aligned grid looking CVT with $16$ centroids in each dimension, with energy $0.1616$, is shown in Fig. \ref{fig:CVT_3D_16x16x16_Gaussian62_51_3p51}.

One of the areas where higher dimensional CVTs have found an application is in the field of evolutionary optimization. Recently introduced, MAP-elites \cite{MAP_elites} is an algorithm that illuminates search spaces in evolutionary optimization, allowing researchers to understand how interesting attributes of solutions combine to affect performance. To scale up the MAP-elites algorithm, authors in \cite{CVT_MAP_elites} employ centroidal Voronoi tessellations, and therein, following \cite{CVTQiangDuParallel}, they employ MacQueen's method to obtain the CVTs and show the sufficiency of using $5000$ centroids. In line with their result, we keep the total number of centroids around the same. Accordingly, we vary $N_i, \forall i \in I_n$ such that $N = \prod_i N_i$ is around $5000$. The corresponding results are given in Table \ref{tab:HighDimensionQiangDu} where we see that the computation time decreases with the increase in dimension $n$. This is because with increase in $n$, we decrease $N_i$ to keep $N$ around $5000$. Hence, the computation of CVT in one-dimensional spaces with fewer centroids is faster. The low energy of all the tessellations is also worth noting. 

\begin{table}[]
\begin{tabular}{@{}cccc@{}}
\toprule
Dimension, $n$ & $N = \prod_i N_i$       & CVT energy            & Time (minutes) \\ \midrule
4              & $4096 = 8^4$            & $0.68 \times 10^{-3}$ & $6.174$                    \\
5              & $4096 = 4^3 \times 8^2$ & $1.2 \times 10^{-3}$  & $4.248$                    \\
8              & $6561 = 3^8 $           & $0.74 \times 10^{-3}$ & $0.641$                    \\
12             & $4096 = 2^{12}$           & $0.21 \times 10^{-3}$ & $0.480$                    \\ \bottomrule
\end{tabular}
\label{tab:HighDimensionQiangDu}
\caption{Tessellation energy and time taken for computing the tessellation for $\Omega_i = [-1,1], \forall i \in I_n$ under density $e^{-10x^2}$ over each $\Omega_1$.}
\end{table}

\section{CONCLUSIONS}
\label{sec::conclusion}

In order to fully rank the quality of all the CVTs under a certain higher dimensional space (read fixed region, number of centroids, and density), the knowledge of all the CVTs and their tessellation energy is required. However, the problem in such a comparison is the difficulty in obtaining CVTs in higher dimension spaces. Nevertheless, the tessellations constructed from CVTs in one-dimensional spaces, that appear to be well drawn out grid-like structures, are proved to be one of the many CVTs in such spaces of dimensions greater than 1. Hence, we are guaranteed to obtain at least one of the many non-unique CVTs in the higher dimensional space. 


Additionally, as seen in the simulation results, the tessellation energy of such CVTs is quantifiably low and are obtained with minimal computational requirement. The method described does not make specific assumptions on the density function, it is scalable to a large number of centroids, and is flexible in terms of dimensions and discretization over each dimension.

However, the (only) key assumption in the construction of CVTs in a higher dimensional spaces from CVTs in one-dimensional spaces is the independence of the densities of the latter spaces. This limits the applicability of the developed idea in areas where CVTs are constrained to a surface, say computing CVT on a sphere, and will be a point of investigation for our future work.

\addtolength{\textheight}{-12cm}   




\bibliography{references}
\bibliographystyle{ieeetr}

\end{document}